\input{epsf}
\documentstyle[preprint,prb,eqsecnum,aps]{revtex}
\tighten
\begin{document}
%
\date{December 14, 1995}
\title{Investigation on the ground states of a model \\
thin film superconductor on a sphere}
\author{M.J.W. Dodgson}
\address{Theoretical Physics Group \\
Department of Physics and Astronomy \\
The University of Manchester, M13 9PL, UK}
\maketitle

\begin{abstract}
We consider the problem of finding the ground state of a model type-II
superconductor on the two-dimensional surface of a sphere, penetrated by $N$ 
vortices. Numerical work shows the ground states to consist of a triangular
 network of the vortices with twelve five-coordinated centres. Values of $N$ 
are found with particularly low energy ground states, due to structures of 
high symmetry. The large $N$ limit is treated within elasticity theory to 
compare with the triangular vortex lattice that forms the ground state on 
an infinite flat plane. Together with numerical work this demonstrates that 
the thermodynamic limit $N\rightarrow\infty$ of the spherical system remains 
different from the flat plane due to the presence of twelve 
disclination defects. 
\end{abstract}

\section{Introduction}

The problem of constructing an optimum lattice-like structure over a 
curved surface 
has become an area of interest in diverse contexts within condensed 
matter physics. Some examples are in work on flexible tethered membranes, 
Fullerene molecules, the Thomson problem of electrons on a spherical surface,
and in models of two-dimensional systems using a spherical geometry to study 
both the Quantum 
Hall Effect (QHE)and thin film superconductors. 

The well known Fullerene molecules 
demonstrate how a low energy structure can be formed by folding a hexagonal 
lattice of carbon atoms (as is found in graphite) onto a closed surface, 
as long as 
twelve five-membered rings (pentagons) are present-- a simple consequence 
of Euler's theorem. These pentagons are essentially disclination defects 
in the hexagonal lattice. The first discovered Fullerene molecule was 
$C_{60}$ in which each of the sixty atoms holds an identical symmetry position 
within the structure of the molecule, so the atoms reside on the surface of 
a sphere.\cite{kroto} This is a special case and as the number of carbon 
atoms increases in these molecules the shape may distort from a sphere to 
reduce the strain 
from the ideal hexagonal lattice over large areas. The interaction between 
the energy cost of bending the surface of the molecule and the strain 
energy within the molecule due to the disclination defects is of central 
importance in the question of the stability of different 
structures.[2--4]

The same considerations are important in the behaviour of membranes with 
internal crystalline order and the ability to buckle out of the 
two-dimensional plane. A disclination defect may lower its energy by 
buckling. If this reduces the total energy of the defect such that it 
diverges at most logarithmically with the system size, then this raises 
the possibility of a buckling transition as defects begin to proliferate 
at some finite temperature.\cite{seung} Alternatively, if we consider a 
closed membrane, such as a vesicle made from surfactant bilayers, 
then the interaction of the internal orientational order 
with the physical curvature may alter the shape,\cite{mack} or even the 
topology\cite{evans} of the membrane.

Some important and well studied models involve these problems but with the 
curvature of the system fixed. One instance is Thomson's problem of trying 
to find the lowest energy configuration for N electrons that are constrained 
to lie on the surface of a sphere.
Although the problem was first proposed in the rather dated context of 
classical models of the atom,\cite{whyte} it has been extensively studied 
recently. This is partly because of the general relevance to any physical 
system of this geometry, but also for its interest as an unsolved problem,
 providing a testing ground for various numerical optimization 
methods.[9--13]

A different model where a two-dimensional system is restricted to a spherical 
surface has been studied both in the context of the QHE and in numerical 
studies on thin film superconductors. In this model, a  magnetic field 
perpendicular to the surface is imposed by placing a Dirac monopole at the 
centre of the sphere. 
The possible electronic states on the surface of the sphere split into Landau 
levels under the influence of the magnetic field, in analogy with the problem 
on a flat plane. The reason the model was put on a sphere by Haldane was to 
allow ``the construction of homogeneous states'' with only a finite number of 
electrons.\cite{haldane} Recent Monte Carlo simulations of 2D superconductors 
have been performed with a  similar model to Haldane's with a 
superconducting wavefunction in the lowest Landau level
on the sphere.[20--22]
This wavefunction  contains $N$ 
zeros that correspond to vortices in the supercurrent, where $N$ depends on 
the quantized strength of the monopole. The reason for using this model is 
again to enable translational invariance which is not possible in a finite 
system on a flat plane.

The Monte Carlo simulations on a sphere have led us to consider in detail 
the ground states of this model. While the problem of finding the ground 
state of an infinite type II 
superconductor penetrated by vortices was long ago solved and found to be 
the triangular vortex lattice,\cite{kleiner} such a lattice cannot form on a 
sphere without the presence of twelve vortices with only five nearest 
neighbours. These twelve vortices are the centres of disclination defects. 
By considering the strains from the perfect triangular lattice, caused by 
the disclination defects, within elasticity theory, we have approximated 
the ground state energy in the large $N$ limit using similar methods  
used in theoretical work on  membranes and Fullerene molecules. 
We have also found numerically 
 the ground states with finite $N$, using symmetry considerations to reach 
 large system sizes. We find  values of $N$ that give particularly 
low energies and this is explained. By extrapolating our numerical results 
to large $N$ we find a finite energy cost per vortex on the sphere 
compared to the infinite plane ground state that is consistent with our 
results from elasticity theory.

\section{Formulation}

Our model thin film superconductor consists of a spherical shell of 
superconducting material, thickness $d$, radius $R$ and a 
monopole at the centre of the sphere that produces an integer multiple of flux 
quanta through the spherical surface. 
We ignore spatial fluctuations in the magnetic flux density, $B$, at the
surface; the effective penetration depth for 
supercurrents in thin films becomes arbitrarily large as $d$ is reduced.
We choose a cylindrically symmetric gauge
consistent with this field with $\hbox{\bf A}\equiv
(A_r,A_\theta,A_\phi)=(0,0,BR\tan{\theta /2})$. 
We measure
lengths in the units $l_m=(\Phi_0/2\pi B)^{1/2}$ which if there are $N$ 
quanta of flux gives $R=(N/2)^{1/2}$.
If we may describe the properties of the superconductor by a complex order 
parameter $\psi(\theta,\phi)$, then the Ginzburg-Landau free energy 
Hamiltonian will be given by
\begin{equation} 
{\cal H}[\psi]=
\int d^3r \left[
\alpha(T) {|\psi|}^2                 
+ \frac{\beta}{2}{|\psi|}^4 
+ \frac{1}{2m} \psi^* {D}^2 \psi \right],
\end{equation}  
where $D^2=\hbox{\bf D}^*.\hbox{\bf{ D}}$ and 
$\hbox{\bf D} = -i\hbar \hbox{\boldmath $\nabla$}
 -2e\hbox{\bf A}$. 
We diagonalize the operator $D^2$ by expanding $\psi$ in a basis of 
eigenfunctions of $D^2$ which form degenerate Landau levels. The degenerate 
set with the lowest eigenvalue may be filled by the orthonormal 
functions\cite{roysing}
\begin{equation} 
\psi_m(\theta,\phi)=
h_m e^{im\phi}
\sin^m ({\theta}/{2})\,
\cos^{N-m} ({\theta}/{2}),
\end{equation}  
with $m=0,N$ and $h_m={[ (N+1)!/4\pi R^2m!(N-m)!]}^{1/2}$.
This is the lowest Landau level (LLL) and over a large range of fields 
and temperatures it is a good approximation to restrict $\psi$ to the LLL, 
$\psi(\theta,\phi)= Q\sum v_m \psi_m(\theta,\phi)$. We set 
$Q=(\Phi_0 k_B T/\beta d B)^{1/4}$. With this restriction 
we can write the Hamiltonian in terms of the  basis coefficients, which 
for $\alpha_T< 0$ is given by
\begin{eqnarray}    
  {\cal H}\left( \alpha_T,\{u_m\} \right)
&\equiv& k_B T \alpha_T^2{\cal F}\left( \{u_m\} \right)\nonumber\\
\label{eq:oneill}
&=k_B T&
\alpha_T^2\left[ -\sum_{m=0}^N u_mu_m^*
+\sum_{p,q,r,s=0}^N w_{p+q,q,r} u_pu_qu_r^*u_s^* \delta_{p+q,r+s} \right],
\end{eqnarray}
where $w_{p+q,q,r}$ is given in \cite{oneill},
$\alpha_T={dQ^2}\left(\alpha(T)+{eB\hbar}/{m}\right)/{k_BT}$ is the 
reduced temperature variable and we have scaled the coefficients as 
$v_m=u_m{|\alpha_T|}^{1/2}$. The quartic term in Equation~(\ref{eq:oneill}) 
can be 
rewritten to give\cite{hanlee}:
\begin{equation}  \label{eq:ham}
{\cal F}\left( \{u_m\}\right) =
\left[ -\sum_{m=0}^N |u_m|^2
+\frac{1}{2N}\sum_{p=0}^{2N} |U_p|^2\right]  ,
\end{equation}  
where $U_p=2\pi N\sum_{q=0}^N B^{1/2}(2N-p+1,p+1)h_qh_{p-q}\Theta(p-q)
\Theta(N+q-p)u_qu_{p-q}$, $B(x,y)=\Gamma(x)\Gamma(y)/\Gamma(x+y)$ is the beta
function and $\Theta(q)$ is the Heaviside step function.

The vortices in this system correspond to the zeros in $\psi(\theta,\phi)$.
The phase of the order parameter changes by $2\pi$ when a path is followed 
that encircles any zero once. This becomes clear if we make the projection
\begin{equation}\label{eq:proj}
\zeta=\tan(\theta/2)e^{i\phi}.
\end{equation}
This gives the form 
$\psi=\cos^N(\theta/2)\sum_{m=0}^N a_m \zeta^m
\equiv\cos^N(\theta/2)f_N(\zeta)$. 
Therefore $f_N(\zeta)$ is a 
holomorphic function of $\zeta$ with $N$ simple zeros in the 
complex-$\zeta$ plane. 
It can always be written in the alternative product form 
$f_N(\zeta)=C\prod_{i=1}^N(\zeta-\zeta_i)$. 
$C$ is an overall complex amplitude and 
$\{\zeta_i\}$ are the vortex positions in the projection 
of Equation~(\ref{eq:proj}).
The function $\psi(\theta,\phi)$ is equally well described by the set 
$\{u_m\}$ or the set $\{C,\zeta_i\}$. There is no simple relation between 
the two, although numerical routines may be used to find the positions 
of the zeros for a given set of basis coefficients. Despite this lack of a 
simple relation from the basis coefficients to coordinates on the sphere, 
there are still some spatial transformations one can make using the 
$\{u_m\}$ formalism. For instance rotation about the $z$-axis by an angle 
$\gamma$ may be performed by the transformation 
$u_m\rightarrow u_me^{im\gamma}$. Reflection in the $x$-$z$ plane results in
$u_m\rightarrow u_m^*$. Rotation by $\pi$ about the $x$-axis occurs under the 
change $u_m\rightarrow u_{N-m}$.

We are interested in the ground states of ${\cal F}\left( \{u_m\}\right)$. 
We write the Hamiltonian as
\begin{equation}
{\cal F}=-\Delta +\frac{\beta_A}{2N}\Delta^2,
\end{equation}
where $\Delta=\sum u_m u_m^*$ and 
$\beta_A=\langle {\left|{\psi_0}\right|}^4\rangle /
{\langle {\left|{\psi_0}\right|}^2\rangle}^2
=\sum |U_p|^2/\Delta^2$ is the Abrikosov factor. This is minimized by 
$\Delta=-N/\beta_A$ to give ${\cal F}_{min}\equiv -NE_0=-N/2\beta_A$, 
so minimizing ${\cal F}$ 
is equivalent to minimizing $\beta_A(\{u_m\})$. The correct $\Delta$ is 
then given by a scale factor on the basis coefficients that does not 
alter $\beta_A$.

The ground state of the LLL vortex system on an infinite plane is well known to
be the triangular lattice\cite{kleiner} which has 
$\beta_A=\beta_{A,0}\simeq 1.1596$. 
With periodic boundary conditions,
this is also the ground state on the finite systems used in other 
simulations.\cite{kato,humac,sasik} 
However, a perfect triangular lattice cannot form on a spherical surface. 
The closest configuration the vortices can make to an
ideal lattice must contain twelve ``disclinations'', i.e.\ twelve vortices that
only have five nearest neighbours.
In Section~\ref{sec:num} we give our results for directly minimizing 
${\cal F}\left( \{u_m\}\right)$ using a simple numerical method, 
but first we 
describe in Section~\ref{sec:elastic} 
calculations using elasticity theory to give the 
finite energy cost that the spherical system will have as 
$N\rightarrow\infty$ due to the twelve disclinations.

\section{Elasticity theory}\label{sec:elastic}

For a lattice in a two-dimensional plane, the elastic energy cost of    
deformations from the perfect ground state lattice is given in the 
harmonic approximation by
\begin{equation}\label{eq:enelastic}
F_{el}=\frac{1}{2}\int d^2r \left(2\mu u_{ij}^2 +\lambda u_{kk}^2\right),
\end{equation}
where $u_{ij}(\hbox{\bf r})=\frac{1}{2}(\partial_iu_{j}(\hbox{\bf r}) 
+ \partial_ju_{i}(\hbox{\bf r}))$ is the elastic strain matrix. 
The displacement \hbox{$\hbox{\bf u}(\hbox{\bf r})=(u_x,u_y)$} represents the 
deformation of the lattice from the point $\hbox{\bf r}$ to the point 
$\hbox{\bf r}+\hbox{\bf u}$. The elastic constants $\mu$ and $\lambda$ are 
related to the shear and bulk moduli by $c_{shear}=\mu$ and 
$c_{bulk}=\mu+\lambda$.
For the LLL ground state the bulk modulus is infinite; the vortex system 
is incompressible so $u_{kk}=0$. The shear modulus is given by\cite{labusch}
\begin{equation}
\mu=0.48\times \frac{1}{2}\mu_0H_{c2}^2\frac{(1-H/H_{c2})^2d}
{2\kappa^2\beta_A^2}.
\end{equation}
The GL parameter $\kappa$ 
is the ratio of magnetic and superconducting correlation 
lengths, which diverges when we neglect the magnetic screening of 
supercurrents.
In our approximation we can write the shear modulus  as 
$\mu=0.48k_BT\alpha_T^2/4\pi \beta_A^2l_m^2=0.0659 E_0/l_m^2$.

From the elastic energy in Equation~(\ref{eq:enelastic}) Hooke's law may be 
derived by 
minimizing $F_{el}$ to give $\partial_i\sigma_{ij}=0$ for the stress tensor
\begin{equation}
\sigma_{ij}=2\mu u_{ij} +\lambda u_{kk}\delta_{ij}.
\end{equation}
The zero divergence condition allows the reformulation of the problem in 
terms of the Airy stress function\cite{nabarro}
$\sigma_{ij}=\epsilon_{ik}\epsilon_{jl}\partial_k\partial_l\chi$.
(This is analogous to the vector potential that ensures zero divergence of 
magnetic fields.)

In the presence of topological defects, such as the 
disclinations we are considering, the displacement field 
$\hbox{\bf u}(\hbox{\bf r})$ is multi-valued. A disclination is defined 
by the change in bond angle 
$\vartheta=\frac{1}{2}\epsilon_{ij}\partial_i u_j$ as a closed loop is 
followed. 
Encircling a five-fold disclination in a triangular lattice will increase 
$\vartheta$ by $2 \pi/6$. This results in the noncommutativity of the 
derivatives of $\vartheta$ at the centre of the disclination. Writing the 
strain field in terms of the Airy stress function results in the biharmonic 
equation that contains all of 2D elasticity theory:\cite{seung}
\begin{equation}\label{eq:biharm}
\frac{1}{K_0}\nabla^4\chi=s(\hbox{\bf r}).
\end{equation}
The density of disclinations is
$s(\hbox{\bf r})=
\sum_\alpha s_\alpha\delta(\hbox{\bf r}-\hbox{\bf r}_\alpha)$
where $\alpha$ labels each defect
and $s_\alpha=2\pi/6$ for a five-fold disclination. 
In Equation~(\ref{eq:biharm}) $K_0$ is the 
2D Young's modulus, which in the LLL is
\begin{equation}
K_0=\frac{4\mu(\mu +\lambda)}{2\mu +\lambda}=4\mu =0.264 E_0/l_m^2.
\end{equation}

Of course, our problem is not on a flat plane but on a sphere, so we 
must take into account the bending of the system out of the plane. In the 
large $N$ limit the surface will be flat locally compared to lattice 
spacings. Over a small region, we may approximate the sphere as a plane with 
some small perpendicular deflection, $f(\hbox{\bf r})$. For our purposes we 
neglect the bending energy which will tend to a constant--- independent on 
the system size--- 
when integrated over the whole sphere. However, we will need to write the 
strain matrix as $u_{ij}=\frac{1}{2}(\partial_iu_{j} 
+ \partial_ju_{i}+\partial_if\partial_jf)$. This alters the biharmonic 
equation by adding an extra term, $\det (\partial_i\partial_j f)\simeq K$, 
the Gaussian curvature to Equation~(\ref{eq:biharm}):
\begin{equation}\label{eq:biharm2}
\frac{1}{K_0}\nabla^4\chi=s(\hbox{\bf r})-K(\hbox{\bf r}).
\end{equation}
For a sphere the curvature is constant, $K=1/R^2$. We write $\chi$ as the 
superposition of twelve contributions corresponding to each disclination,
$\chi(\hbox{\bf r})=\sum_{\alpha=1}^{12}\chi_{\alpha}(\hbox{\bf r})$. 
The solution to Equation~(\ref{eq:biharm2}) is found in
\ref{app:airy}. The elastic energy from Equation~(\ref{eq:enelastic}) 
can be
 written in terms of the Airy stress function as in Equation~(\ref{eq:enfin}). 
We have calculated the energy cost for these twelve disclinations for 
different configurations.  For any configuration the total energy scales with 
the surface area of the sphere, so there is a finite energy cost per 
vortex in the limit $N\rightarrow\infty$. 
The minimum energy is found with the disclinations 
at the corners of an icosahedron. In this case we find an energy cost 
per vortex of
\begin{equation}\label{eq:eldiff}
\delta E_{el}=0.047 K_0R^2/N=0.0031 E_0.
\end{equation}
In Section~\ref{sec:num} we will compare $\delta E_{el}$ with 
the energy $\delta E_{el}$ obtained by direct minimization of 
${\cal F}(\{u_m\})$ for finite $N$.

\section{Numerical results}\label{sec:num}

\begin{figure}[htbp]           
\epsfxsize= 8cm
\begin{center}
\leavevmode\epsfbox{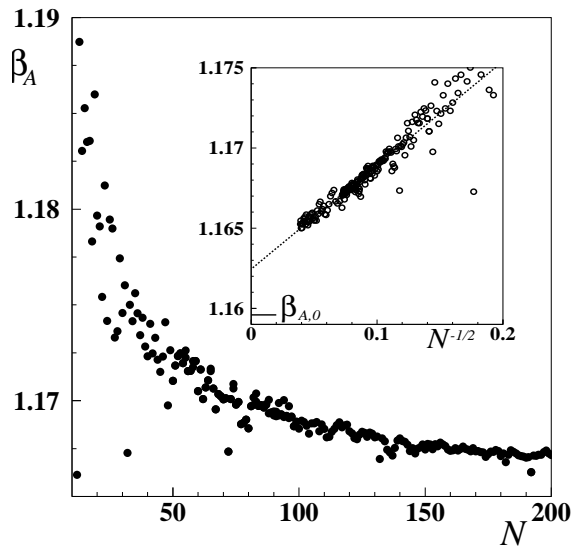}
\\         
\caption{
The minimum values of the Abrikosov ratio $\beta_A$
for different system sizes. 
Note the low values for $N=12,32,72,132$ and $192$.
The 
inset shows the values for large $N$ plotted against $N^{-1/2}$ with a fit 
to extrapolate to the $N\rightarrow\infty$ limit. The value for an 
infinite flat plane, $\beta_A,0$, is shown for comparison.
 \label{fig:1}}
\end{center}\end{figure}

We have used a simple quasi-Newton algorithm to find 
configurations $\{u_n\}$ that 
minimize the Abrikosov ratio $\beta_A(\{u_n\})$; this is equivalent to 
finding the minimum energy of the system. Clearly there are some 
transformations of $\{u_n\}$ under which $\beta_A$ is invariant. The energy 
of the system remains unchanged after a global phase change in the order 
parameter, or after a rotation of the whole system about some axis. These 
freedoms
can be fixed by  restrictions on the coefficients.\cite{oneill} 
Our results up to $N=200$ are shown in Figure~\ref{fig:1}. 
The presence of ``magic numbers'', for which the ground 
state has a  lower energy than nearby values, is clearly seen at
$N=12$, $32$, $72$, $132$ and $192$. To some extent we can explain the magic 
numbers from the expected symmetry 
of the most stable ground states.

\begin{figure}[htbp]           
\epsfxsize= 15cm
\begin{center}
\leavevmode\epsfbox{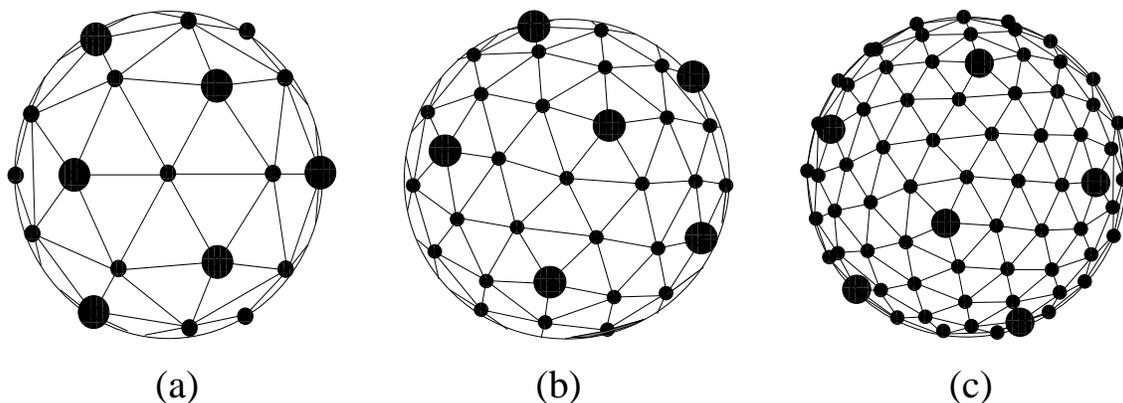}
\\      
\caption{ The numerically found ground states for three magic number cases:
(a) $N=32$, (b) $N=72$, (c) $N=132$. Only one hemisphere is shown; the dots 
represent the zeros of the order parameter and the five-fold centres are 
represented by larger dots, which in each case form the corners of an
icosahedron.
\label{fig:2}}
\end{center}\end{figure}

By finding the zeros of the ground states, we can look at the vortex
configurations. As might be expected they make up a triangular network, but
with twelve five-coordinated centres. The magic number states display 
icosahedral symmetry with the disclinations corresponding to the corners of 
the icosahedron, as in Figure~\ref{fig:2}. In fact the structures appear to 
be projections on to the sphere of icosadeltahedra, which are polyhedra with 
identical equilateral triangular faces and icosahedral symmetry.\cite{caspar}
These may be constructed by considering the number of triangular lattice 
vectors between neighbouring five-fold centres, labelled by the indices 
$(h,k)$. 

A simple geometrical argument shows that this divides each face of the 
icosahedron into $T$ triangular faces with $T=h^2+hk+k^2$, so T may be equal 
to $1,3,4,7,9,12,13,16,19 ...$. This gives a total of $20T$ faces and $10T+2$ 
vertices to the icosadeltahedra.  The magic numbers we find satisfy 
these conditions, the only remaining question being why some possible values 
of $N$ are not so low in energy, eg.\ $N=42$ or $122$.

Notice that the structures for $N=72$ and $132$ possess a chirality. In fact 
this will always be the case for $h\ne k$ (as long as neither is zero). 
This may be a factor in deciding the lowest energy structures as 
the complex conjugation of the coefficients is equivalent to a reflection. 
If a ground state has no chirality then any reflection of the state will be 
equivalent to a rotation. This would require that we could write the ground 
state  with all coefficients having the same phase. It is unlikely that this 
combination would be effective in minimizing  the complex interactions in the 
quartic term of the Hamiltonian especially as the system size increases. 
Therefore the chirality allows a greater variety of phase differences 
between the coefficients. Such subtleties may explain why the ground 
states for some non-chiral icosahedral numbers are not particularly stable,
as with $N=42,122$.

Remarkably, the same structures that are found when we minimize our vortex 
system for some magic numbers are seen in nature in the form of the shells of 
certain viruses.\cite{klug} In particular the structure shown in 
Figure~\ref{fig:2}(c) for $N=132$ is also observed in double and single 
shelled simian rotaviruses, in a left handed configuration.\cite{prasad} 
That these similarities exist in such different systems suggests that some 
general principle exists for the criterion of the most stable structure. 
It is possibly related to a mathematical problem that also generates these 
structures. This is the ``covering'' problem: How may $N$ equal overlapping 
circles (without gaps) cover a sphere so that the diameter of the circles is 
minimized?\cite{tarnai} The centres of these circles correspond to the 
vortices in our system. The alternative problem of maximizing the diameter 
of the circles apparently gives different solutions.

The details of the magic number states becomes less important as $N$ becomes 
 large, the limit treated in Section~\ref{sec:elastic} within elasticity 
theory.
As $N$ increases the numerical minimization becomes less trivial as there 
is an increasing overlap between the
basis states resulting in more complex phase interference (see 
Ref.~\onlinecite{mosch} where
numerical minimization on a plane in cylindrical coordinates resulted in rather
high values of $\beta_A$). Another possibly related problem is the growth in
 the number of metastable states with energies only slightly above the ground 
state. The numerical work on Thomson's problem has found that the number of 
metastable states grows exponentially with the system size.\cite{erber}
We may use our knowledge of the symmetry of the most stable configurations 
to  reduce the number of free variables by an order of magnitude. We assume 
the icosahedral properties of the ground state, and choose a five-coordinated 
vortex at $\theta=0$. The five-fold symmetry about the $z$-axis is imposed by 
setting $u_n=0$ for $n\ne (5m+1)$ where $m$ is any integer. 
The icosahedron also has five two-fold axes 
of symmetry at right angles to each five-fold axis. This two-fold rotational 
symmetry will arise if we set $u_n=u_{N-n}$. We have performed the same 
minimization routine using these constraints for $N=10m+2$ up to $N=652$.

Our results for large $N$ are shown in the inset to Figure~\ref{fig:1}.  
The data fits 
well to the form $\beta_A = A+BN^{-1/2}$ with 
$A\simeq 1.1624=\beta_{A0}(1+0.0024)$ and $B\simeq 0.0648$.
As $N\rightarrow\infty$, $\beta_A$ does not seem to converge to the infinite 
plane value $\beta_{A0}$ (contrary to the conclusions of 
Ref.~\onlinecite{oneill}
from the minimizations of small system sizes). 
This extrapolation implies a finite energy cost per vortex on the sphere in 
the 
large $N$ limit of
\begin{equation}
\delta E=0.0024 E_0.
\end{equation}
The difference between this and the result of Equation~(\ref{eq:eldiff}) 
from elasticity 
theory may be explained by the inadequacy of the harmonic approximation in 
this calculation. As there are large strains associated with the disclination 
defects, non-linear effects will be important in determining the total energy 
cost. In their calculation of disclination defects on flat membranes 
Nelson and Seung\cite{seung} found the same mismatch between elasticity 
theory and 
numerical results, with elasticity overestimating the energy cost by a 
similar proportion.

Within the approximations of Section~\ref{sec:elastic} the correction for 
finite $N$ is not predicted. Numerically the energy cost per vortex falls 
as $1/R$ at large $N$ which implies a total contribution that grows 
proportionally to the radius. 

It must be stressed that the results of this section required
no great numerical effort. More sophisticated optimization methods 
(eg.\ simulated annealing\cite{wille} and its generalizations\cite{alt}) 
may give greater 
confidence in whether or not the absolute minima have been found. 
More extensive 
work would also give results for larger $N$.
However, our use of symmetry has allowed us to do a great deal with just a 
simple routine.

\section{Conclusions}

The original motivation of this work was for the use of the numerical ground 
states in Monte Carlo simulations.\cite{dodgson} 
We also wanted to investigate the 
differences between these ground states and the ground states on a flat plane,
for which other groups have performed simulations obtaining different results. 
The work in this paper shows that the vortex ground states on the sphere do
not 
approach the ground state of the infinite flat plane as 
$N\rightarrow\infty$. The presence of the twelve disclinations remains 
important however large the sphere.

This work may also be of interest in wider contexts: first, in its relation 
to other optimization problems of points on a sphere. Our particular system 
allows a use of symmetry that may not be so straightforward with position 
variables. This enables us to find approximate ground states at  large 
$N$ with quite unsophisticated numerical techniques. Our elasticity 
calculation may be relevant to the large $N$ limit of Thomson's problem, 
using properties of the Wigner lattice on a 2D infinite plane. This limit 
has been considered before, and projection of the Wigner lattice onto a 
spherical surface was used to estimate the extra energy on the 
sphere.\cite{glasser} However, no consideration was taken of the 
required disclination defects. This paper also provides a numerical test 
of the accuracy of elasticity theory for curved membranes and disclination 
defects where the approximation of small deviations from the ideal lattice 
breaks down. Finally, the fact that the structures we see in the magic 
number ground states are the same structures seen in such a different 
field as virology suggests that these fascinating shapes are the result 
of some  general optimization criteria.

\acknowledgements

 It is a pleasure to thank Mike Moore for his help and encouragement as 
well as Mike Evans for useful conversations.

\appendix
\section{Disclination on a sphere}\label{app:airy}

In this appendix we derive the contributions from each of the
twelve disclinations to the the Airy Stress function 
$\chi(\hbox{\bf r})=\sum_{\alpha=1}^{12}\chi_{\alpha}(\hbox{\bf r})$ 
and describe how this leads to the elastic energy cost of these disclinations.
From Equation~(\ref{eq:biharm2}) with $K=1/R^2$ and 
$s=\frac{\pi}{3}\sum_{\alpha=1}^{12}\delta^2(\hbox{\bf r}-
\hbox{\bf r}_{\alpha})$
we have:
\begin{equation}\label{eq:biharm3}
\frac{1}{K_0}\nabla^4\chi_{\alpha}=\frac{\pi}{3}\delta^2(\hbox{\bf r}-
\hbox{\bf r}_{\alpha})-\frac{1}{12R^2},
\end{equation}
where $\hbox{\bf r}-\hbox{\bf r}_{\alpha}=
(\theta'(\alpha),\phi'(\alpha))$, and $\theta'(\alpha)$ and 
$\phi'(\alpha)$ are the polar and azimuthal angles with respect to the axis 
through the  disclination.
Using the symmetry about this axis, this may be integrated to give
\begin{equation}\label{eq:del2}
\nabla^2\chi_{\alpha}=\frac{K_0}{12}(\ln \sin{(\theta'(\alpha)/2)}+A),
\end{equation}
with $A$ a constant. For $\chi_{\alpha}$ to be well defined the 
integral of $\nabla^2\chi_{\alpha}$ over the whole sphere must be zero, 
which means
$A=\ln{2}-1$. Integrating again leads to
\begin{eqnarray}
 \frac{\partial\chi_{\alpha}}{\partial\theta'(\alpha)}=
\frac{K_0R^2}{24}\tan{(\theta'(\alpha)/2)}&&\left(
2\ln{\sin{(\theta'(\alpha)/2)}}+3(\ln{2}-1)\right)\\
 \chi_{\alpha}=\frac{K_0R^2}{12}\{
-\ln{\cos{(\theta'(\alpha)/2)}}&&\left[2\ln{\sin{(\theta'(\alpha)/2)}}
+3(\ln{2}-1)\right]\nonumber\\
&&+\int_0^{\sin{\frac{\theta'(\alpha)}{2}}}
\frac{\ln{(1-x^2)}}{x}dx\}.\label{eq:imp}
\end{eqnarray}
The integral in Equation~(\ref{eq:imp}) cannot be written as a finite number 
of 
elementary functions.\cite{ryzhik}
From Equation~(\ref{eq:enelastic}) the elastic energy of the disclinations 
can be 
written
\begin{eqnarray}
F_{el}&=&\frac{1}{2}\int d^2r \left[\frac{1+\sigma}{K_0}
( \partial_i\partial_j\chi)^2
-\frac{\sigma}{K_0}( \nabla^2\chi)^2\right]\\
&=&\frac{1}{2}\int d^2r \left[\frac{1}{K_0}( \nabla^2\chi)^2
-\frac{1+\sigma}{K_0}\epsilon_{ik}\epsilon_{jl}\partial_k\partial_l
\left( \partial_i\chi\partial_j\chi\right)
\right],\label{eq:enfin}
\end{eqnarray}
with the 2D poisson ratio $\sigma=\lambda/(2\mu+\lambda)$ equal to unity 
in the LLL approximation.
The second term in Equation~(\ref{eq:enfin}) only gives contributions on 
boundaries,
and so is zero on the sphere.
Therefore from Equation~(\ref{eq:del2}) we can find the energies of different 
configurations of the disclinations on the sphere.

\newpage


\begin{thebibliography}{999}

\bibitem{kroto}  Kroto H W 1992 {\it Angew.\ Chem.} {\bf 31} 111 
\bibitem{tersoff}  Tersoff J 1992  {\it Phys. Rev.} B {\bf 46} 15546 
\bibitem{witten} Witten T A  and Li H 1993 {\it Europhys.\ Lett.} 
{\bf 23} 51
\bibitem{zhang}  Zhang Z, Davis H T, Maier R S and Kroll D M 1995 
 {\it Phys. Rev.} B {\bf 52} 5404 
\bibitem{seung} Seung H S and  Nelson D R 1988  {\it Phys. Rev. A} 
{\bf 38} 1005 
\bibitem{mack} MacKintosh F C and  Lubensky T C 1991  {\it Phys. Rev. Lett.} 
{\bf 67} 1169;  Park J, Lubensky T and  MacKintosh F C 1992
{\it Europhys.\  Lett.} {\bf 20} 279  
\bibitem{evans} Evans R M L {\it J.Physique} 1995 {\bf 5} 507 
\bibitem{whyte} Whyte L L 1952 {\it Am.\ Math.\ Monthly} {\bf 59}, 606
\bibitem{wille} Wille L T {\it Nature} 1986 {\bf 324} 46 
\bibitem{erber} Erber T and  Hockney G M 1991 {\it J. Phys. A: Math. Gen.} 
{\bf 24} L1369; 
Erber T and  Hockney G M 1995  {\it Phys. Rev. Lett.} {\bf 74} 1482 
\bibitem{glasser} Glasser  L and Every A G 1992 {\it J. Phys. A: Math. Gen.} 
{\bf 25} 2473
\bibitem{edmund} Edmundson J R 1993 {\it Acta Crystallogr.} A {\bf 49} 648 
\bibitem{alt} Altschuler E L, Williams T J, Ratner E R, 
 Dowla F and Wooten F 1994  {\it Phys. Rev. Lett.} {\bf 72} 2671 
\bibitem{haldane} F. D. M. Haldane F D M 1983  {\it Phys. Rev. Lett.}
 {\bf 51} 605 
\bibitem{oneill} O'Neill J A and  Moore M A 1992  {\it Phys. Rev. Lett.}
 {\bf 69} 
2582;  O'Neill J A and  Moore M A 1993  {\it Phys. Rev.} B {\bf 48} 374  
\bibitem{hanlee} Lee H H and  Moore M A  {\it Phys. Rev.}  B {\bf 49}9240 
\bibitem{dodgson} Dodgson M J W and  Moore M A (submitted to  
{\it Phys. Rev.} B).
\bibitem{kleiner} Kleiner W H, Roth L M and  Autler S H 1964  {\it Phys. Rev.}
 {\bf 133} 1226
\bibitem{roysing} Roy  S M and Singh V 1983   {\it Phys. Rev. Lett.}
 {\bf 51} 2069  
\bibitem{kato} Kato Y and  Nagaosa N 1993  {\it Phys. Rev.} B {\bf 47}2932; 
Kato Y and  Nagaosa N 1993  {\it Phys. Rev. B} {\bf 48} 7383 
\bibitem{humac}  Hu J and MacDonald A H 1993 {\it  Phys. Rev. Lett.}
 {\bf 71} 432; 
Hu J and MacDonald A H 1994  {\it Phys. Rev. B} {\bf 49} 15263
\bibitem{sasik}  Sasik R and  Stroud D 1993  {\it Phys. Rev. B} {\bf 49} 16074;
 Sasik R,  Stroud D and  Tesanovic  Z 1995  {\it Phys. Rev. B} {\bf 51} 3042  
\bibitem{labusch} Labusch R 1969 {\it Phys. Status Solidi}  {\bf 32} 439 
\bibitem{nabarro} Nabarro F R N 1967 {\it Theory of Crystal Dislocations}
(Oxford: Clarendon Press)
\bibitem{caspar}  Caspar D L D 1993 {\it Phil.Trans. R. Soc.} A {\bf 343} 133 
\bibitem{klug}  Caspar D L D and Klug A 1962
 {\it Cold Spring Harbor Symp.\ Quant.\ Biol.} {\bf 27} 1 
\bibitem{prasad} Prasad B V,  Wang G J,  M. Clerx J P M and Chiu W 1988
{\it J.\ Mol.\ Biol.} {\bf 199} 269 
\bibitem{tarnai} Tarnai T, {\it J.\ Mol.\ Biol.} 1991 {\bf 218} 485;
Tarnai T and  Gaspar Zs 1991 {\it Math.\ Proc.\ Camb.\ Phil.\ Soc.} 
{\bf 110} 71 
\bibitem{mosch} Moshchalkov V V, Dhalle M and  Bruynseraede Y 1993
{\it Physica} C {\bf 207} 307 
\bibitem{ryzhik} Gradshteyn I S and Ryzhik I M 1965 
{\it Table of Integrals, Series, and Products} (New York: Academic Press), 
p. 205.
\end{thebibliography}
\end{document}